\documentclass[aps,prx,10pt,showpacs,amssymb,nofootinbib,superscriptaddress,onecolumn,longbibliography]{revtex4-2}
\usepackage{subcaption}
\usepackage[margin=8pt,font=small,labelfont=bf]{caption}
\usepackage{physics}
\usepackage{tikz}
\usetikzlibrary{shapes,snakes,backgrounds,fit,decorations.pathreplacing,arrows,decorations.markings}
\usepackage{verbatim}
\tikzstyle{vecArrow} = [thick, decoration={markings,mark=at position
   1 with {\arrow[semithick]{open triangle 60}}},
   double distance=1.4pt, shorten >= 5.5pt,
   preaction = {decorate},
   postaction = {draw,line width=1.4pt, white,shorten >= 4.5pt}]
\tikzstyle{innerWhite} = [semithick, white,line width=1.4pt, shorten >= 4.5pt]

\usepackage[ansinew]{inputenc}
\usepackage{mathtools} 
\usepackage{bbm}
\usepackage{bm}
\usepackage{amsbsy}
\usepackage{amsthm}
\usepackage{amssymb}
\usepackage{amsfonts}
\usepackage{upgreek}
\usepackage{amsmath}
\usepackage{dsfont} 
\usepackage{graphicx} 
\usepackage{epsfig}
\usepackage{epstopdf}
\usepackage{dsfont}
\usepackage{multibib}
\usepackage{color}

\usepackage[colorlinks]{hyperref}
\makeatletter
\newcommand\org@hypertarget{}
\let\org@hypertarget\hypertarget
\renewcommand\hypertarget[2]{%
  \Hy@raisedlink{\org@hypertarget{#1}{}}#2%
  }
\makeatother
\usepackage[figure,table]{hypcap}
\usepackage{MnSymbol}
\usepackage{enumerate}
\usepackage{ragged2e}
\usepackage{float}
\usepackage{orcidlink}
\hypersetup{
	bookmarksnumbered,
	pdfstartview={FitH},
	citecolor={darkgreen},
	linkcolor={darkred},
	urlcolor={darkblue},
	pdfpagemode={UseOutlines}}
\definecolor{darkgreen}{RGB}{50,190,50}
\definecolor{darkblue}{RGB}{0,0,190}
\definecolor{darkred}{RGB}{238,0,0}
\definecolor{quantum}{RGB}{83,37,127}
\definecolor{quantumlight}{RGB}{169,146,191}
\usepackage{soul}

\newcommand{\deff}{d\ensuremath{_{\hspace{-0.5pt}\protect\raisebox{0pt}{\tiny{eff}}}}}

\renewcommand{\thesection}{\arabic{section}}

\makeatletter
\renewcommand{\p@subsection}{}
\renewcommand{\p@subsubsection}{}
\makeatother

\newcommand{\M}[1]{\mathcal{#1}}

\newcommand{\up}{\texttt{Up} }
\newcommand{\dw}{\texttt{Dw} }
\newcommand{\mypm}{\texttt{Pm} }

\newtheorem{Theorem}{Theorem}
\newtheorem{Lemma}{Lemma}
\newtheorem{Definition}{Definition}

\usepackage{paracol}
\usepackage{tcolorbox}

\begin{document}

\title{A Complexity-Based Approach to Quantum Observable Equilibration}

\author{Marcos G. Alpino\,\texorpdfstring{\orcidlink{0009-0000-2390-4086}}{}}
\affiliation{Departamento de F\'{\i}sica - ICEx - Universidade Federal de Minas Gerais,
Av.~Pres.~Ant\^{o}nio Carlos 6627 - Belo Horizonte - MG - 31270-901 - Brazil.}
\author{Tiago Debarba\,\texorpdfstring{\orcidlink{0000-0001-6411-3723}}{}}
\affiliation{Departamento Acad{\^ e}mico de Ci{\^ e}ncias da Natureza - Universidade Tecnol{\'o}gica Federal do Paran{\'a}, Campus Corn{\'e}lio Proc{\'o}pio - Paran{\'a} -  86300-000 - Brazil.}
\author{Reinaldo O. Vianna\,\texorpdfstring{\orcidlink{0000-0003-2857-8552}}{}}
\affiliation{Departamento de F\'{\i}sica - ICEx - Universidade Federal de Minas Gerais,
Av.~Pres.~Ant\^{o}nio Carlos 6627 - Belo Horizonte - MG - 31270-901 - Brazil.}

\author{Andr\'e T. Ces\'{a}rio\,\texorpdfstring{\orcidlink{0000-0002-6972-2576}}{}}
\email{andretcs@ufmg.br}
\affiliation{Departamento de F\'{\i}sica - ICEx - Universidade Federal de Minas Gerais,
Av.~Pres.~Ant\^{o}nio Carlos 6627 - Belo Horizonte - MG - 31270-901 - Brazil.}

\begin{abstract}
We investigate the role of a statistical complexity measure to assign equilibration in isolated quantum systems. While unitary dynamics preserve global purity, expectation values of observables often exhibit equilibration-like behavior, raising the question of whether a measure of complexity can track this process. In addition to examining observable equilibration, we extend our analysis to study how the complexity of the quantum states evolves, providing insight into the transition from initial coherence to equilibrium. We define a classical statistical complexity measure based on observable entropy and deviation from equilibrium, which captures the dynamical progression towards equilibration and effectively distinguishes between complex and non-complex trajectories. In particular, our measure is sensitive to non-complex dynamics. Such dynamics include the quasi-periodic behavior exhibited by low-dimensional initial states, where the system explores a limited region of Hilbert space while preserving coherence. Numerical simulations of an Ising-like non-integrable Hamiltonian spin-chain model support these findings. Our work provides new insight into the emergence of equilibrium behavior from unitary dynamics and advances complexity as a meaningful tool in the study of the emergence of classicality in microscopic systems.
\end{abstract}

\maketitle



\section{Introduction}
Understanding equilibration in quantum systems - how a system evolves from an initial pure state to an apparent equilibrium - is a central problem in the foundations of quantum mechanics. Traditionally, this process is linked to the system reaching a state of maximal disorder or entropy. But what happens to the \emph{complexity} of the quantum state during this process? This work explores whether a measure of complexity can serve as a meaningful quantifier of equilibration, with a focus on systems defined by a Hamiltonian $H$, an observable $O$, and a simple, zero-entropy initial state $\ket{\psi_0}$. We ask: Can we use a complexity measure to quantify \emph{how much} a system has equilibrated? This question drives our investigation into the role of a complexity measure in observable equilibration processes.

The foundational question of how macroscopic irreversibility emerges from time-symmetric quantum dynamics dates back to Boltzmann and has shaped the development of statistical mechanics \cite{Landau1980, lebowitz1993macroscopic, zwanzig1970concept, brown2009boltzmann, steckline1983zermelo}. In modern quantum theory, the puzzle reemerges in the form of understanding how closed quantum systems equilibrate \cite{gogolin2016equilibration,Malabarba16}. Recent studies have shown that, despite unitary evolution, expectation values of observables can relax to long-time averages that exhibit an observable-dependent thermal equilibrium \cite{linden2009quantum, short2011equilibration,reimann2018dynamical}. This has led to the concept of \emph{Observable Equilibration}, which emphasizes the classical statistical behavior of measurement outcomes rather than the full quantum state \cite{schwarzhans2023quantum,engineer2024equilibration,demelo2024finite}. Recent results further support this perspective by showing that the emergence of a second law in isolated quantum systems can be captured through statistical properties of observables \cite{meier2025emergence}.
 Furthermore, observable equilibrium states are, on average, diagonal in the Hamiltonian eigenbasis, lacking coherence  \cite{Scarpa2023, Anza2017}. This observation invites a resource-theoretic interpretation: on average, coherence, like free energy, becomes a resource consumed in the equilibration process \cite{lostaglio2015description,gour2022role}. This link aligns with the association of thermodynamic irreversibility with coherence depletion \cite{crooks1998nonequilibrium}. In this work, we revisit this concept through the lens of statistical complexity. The classical statistical complexity measure introduced by L\'opez-Ruiz \emph{et al.} quantifies structure in probability distributions by combining entropy and deviation from microcanonical state \cite{lopez1995}. This idea has been extended to quantum systems through the Quantum Statistical Complexity Measure $(QSCM)$, which signals transitions between ordered and disordered quantum states \cite{cesario2022quantum}.

We propose a complexity-based approach to study equilibration that incorporates the statistical structure of observable outcomes. We aim to understand whether a measure of complexity can indicate that a system has equilibrated, and whether it can capture the subtle transition from quantum coherence to classical equilibrium. The formalism of observable equilibration builds on this by examining the long-time behavior of expectation values and probability distributions associated with physical observables. These distributions can exhibit relaxation, transient oscillations, and effective stabilization features, suggesting a rich structure in the system's evolution.

These features are studied adopting a probabilistic and operational perspective. While the global quantum state remains pure throughout unitary evolution, the observable statistics, particularly those tied to physical measurements like total magnetization \emph{per} spin, \textit{etc.}, reveal information-theoretic signatures of equilibration. Based on the results obtained in Ref.~\cite{meier2025emergence}, we propose a Classical Complexity Statistical Measure, named as \textit{Observable Equilibration Complexity measure} (OECM), built from observable entropy and distance to equilibrium distributions. This leads to the notion of the observable equilibration complexity measure, which quantifies the state's both structural and temporal informational contents concerning a chosen observable. 

Numerical analysis confirms that the proposed Observable Equilibration Complexity Measure effectively captures the system's equilibration behavior. Our simulations on a non-integrable Ising spin chain of $N = 10$ spin-$1/2$ particles were initialized in three distinct pure quantum states, each exhibiting different dynamical regimes based on the effective dimension of the initial state: the fully polarized up~state, $\ket{\uparrow \uparrow \dots \uparrow}$ (\texttt{Up}), the fully polarized down state, $\ket{\downarrow \downarrow \dots \downarrow}$ (\texttt{Dw}), and the alternating paramagnetic configuration, $\ket{\uparrow \downarrow \uparrow \downarrow \dots}$ (\texttt{Pm}). These configurations allow us to explore a range of equilibration scenarios across different effective dimensions.

For initial configurations with higher effective dimensions, such as \dw and \mypm states, we observe a gradual and sustained decay of the Observable Equilibration Complexity Measure towards zero, in agreement with theoretical predictions. This behavior reflects the system's enhanced capacity to explore a larger portion of the Hilbert space and, therefore, it facilitates equilibration. Conversely, the \up state, characterized by a significantly lower effective dimension, exhibits quasi-periodic dynamics with limited delocalization across the energy eigenbasis. As a result, the Observable Equilibration Complexity Measure displays a comparatively faster decay, indicative of a less complex trajectory. These numerical results not only corroborate the analytical bounds derived in this work but also underscore the utility of statistical complexity measures as a diagnostic tool for distinguishing between complex equilibration dynamics and simpler, coherence-preserving evolutions that characterize transitions from quantum initial coherence states to classical-like equilibrium behavior.

The paper is structured as follows. In Section~\ref{sec:framework}, we present the mathematical setup and define equilibrium in terms of dephased states. Section~\ref{sec:complexity_bounds} introduces the statistical complexity measures, defines the Observable Equilibration Complexity Measure, and provides bounds on their evolution during equilibration. Section~\ref{sec:numerics} evinces these concepts numerically using a non-integrable Ising-like spin-chain model, and Section~\ref{sec:conclusion} offers final remarks and open questions. Finally, Appendix~\ref{app_A} complements the discussion by providing a numerical evaluation of the theoretical upper bounds for the observable complexity measures, offering further insight into their behavior and limitations in practical scenarios.

\section{Framework}\label{sec:framework}

We consider a finite-dimensional quantum system of dimension $d$, governed by a Hamiltonian $H \in \M{B}(\M{H})$ with spectral decomposition
\begin{equation}
    H = \sum_{i=1}^{n} E_i \Pi_i,
\end{equation}
where $n$ is the number of distinct eigenvalues (with $n \leq d$), and $\tr(\Pi_i) = d_i$ corresponds to the degeneracy of energy level $E_i$. The total system dimension satisfies $\sum_{i=1}^{n} d_i = d$. The system evolves according to the unitary dynamics generated by $H$, described by
\begin{equation}
    U_t = e^{-i H t},\text{ with } t \in \mathbb{R}.
\end{equation}
Given an initial state $\rho_0$, the evolved state at time $t$ follows
\begin{equation}
    \rho_t = U_t \rho_0 U_t^\dagger,
\end{equation}
which solves the Schr\"{o}dinger equation $\dot{\rho}_t = -i[H, \rho_t]$. The expectation values of observables $O \in \M{B}(\M{H})$, decomposed as $O = \sum_{l=1}^r o_l P_l$, with $r$ the rank of $O$, evolve as
\begin{equation}
   \label{eq:observable}
    \tr(O \rho_t) = \tr(O U_t \rho_0 U_t^\dagger) = \tr(U_t^\dagger O U_t \rho_0).
\end{equation}
The latter form of Eq.~\eqref{eq:observable} reveals the evolution in the Heisenberg picture. An important quantity in the study of quantum equilibration is the \emph{effective dimension} of the initial state concerning the Hamiltonian statistics. This quantity is defined as 
\begin{equation}
\label{eq:d_eff}
d_{\mathrm{eff}} = \left( \sum_i \mathrm{tr}(\Pi_i \rho_0)^2 \right)^{-1}\!.
\end{equation}

The effective dimension, \( d_{\text{eff}} \), is a measure of how the initial state \(\rho_0\) is spread across the eigenstates of the Hamiltonian. Specifically, it quantifies the degree to which the state is delocalized in the energy eigenbasis of the system. A high value of \( d_{\text{eff}} \) implies that the initial state occupies many energy levels, whereas a low value indicates that the initial state is concentrated in a smaller number of energy eigenstates. This quantity is particularly relevant when analyzing the approach to equilibrium, as systems with a large effective dimension tend to exhibit faster relaxation to equilibrium due to the greater number of accessible states. Conversely, systems with a small effective dimension may exhibit slower equilibration, as fewer energy levels are involved in the evolution \cite{linden2009quantum}. For a function \( g(t) \) defined over a finite time interval \( [0, T] \), we define the time average of \( g(t) \) as
\begin{equation}\label{eq:observable_timeaverage}
    \langle g \rangle_T = \frac{1}{T} \int\limits_0^T g(t) \, dt,
\end{equation}
which represents the average value of the function over the time interval \( [0, T] \). This quantity is useful for quantifying the behavior of a system over a finite period, providing an estimate of the long-term behavior for systems that exhibit periodic or transient dynamics. The infinite-time average is defined as the limit of the time average as \( T \to \infty \), given by
\begin{equation}
   g_\infty =  \lim_{T \to \infty} \langle g \rangle_T.
\end{equation}

This quantity describes the steady-state behavior of the system, where \( g(t) \) approaches a constant value as time progresses. The infinite-time average is particularly important when studying equilibrium states, as it represents the asymptotic value that observables reach after a sufficiently long time, assuming the system has equilibrated.

\subsection{Equilibration of Observables}

In isolated quantum systems, equilibration refers to the process in which the expectation values of observables stabilize at long-time averages. This occurs due to the unitary evolution of the system, and the dynamics is influenced by the triple $(H, \rho_0, O)$, where $H$ is a non-integrable Hamiltonian, $\rho_0$ is the initial state, and $O$ is the observable.

The equilibrium state, denoted by $\omega$, represents the long-time average state of the system. It is obtained by taking the time integral of the system's state $\rho_t$ over the interval $[0, T]$ and then letting~$T \to \infty$. Mathematically, the equilibrium state is expressed as
\begin{equation}
    \omega = \lim_{T \to \infty} \frac{1}{T} \int_0^T \rho_t \, dt.
\end{equation}

It can be shown \cite{short2011equilibration} that the equilibrium state $\omega$ is the dephased version of the initial state $\rho_0$ in the Hamiltonian eigenbasis. This means that $\omega$ is a diagonal matrix in the eigenbasis of the Hamiltonian, and it can be written as

\begin{equation}
    \omega = \sum_i \Pi_i \rho_0 \Pi_i,
\end{equation}
where $\Pi_i$ are the projectors onto the eigenstates of the Hamiltonian. This dephasing process is a key feature of equilibration, as it effectively removes any off-diagonal coherence in the energy eigenbasis, thus leading to a state where all observable quantities are stationary.

The concept of \emph{effective dimension} is intimately related to the equilibrium state. The effective dimension \( d_{\text{eff}} \) quantifies how widely the initial state \( \rho_0 \) is distributed over the energy eigenstates of the Hamiltonian. It can be equivalently expressed as $d_{\text{eff}} = (\tr(\omega^2))^{-1}.$

According to Reimann and Kastner \cite{Reimann2012}, under suitable conditions, the time-averaged deviation of an observable from its equilibrium expectation is bounded by
\begin{equation}\label{eq:reimann_bound}
    \langle |\tr(O \rho_t) - \tr(O \omega)|^2 \rangle_T \leq \frac{\|O\|^2}{d_{\text{eff}}} f(\epsilon, T),
\end{equation}
where $\|O\|$ denotes the usual operator norm, and $f(\epsilon, T)$ captures properties of the energy spectrum
\begin{equation}\label{eq:fepsilonT}
    f(\epsilon, T) = N(\epsilon) \left(1 + \frac{8 \log_2(n)}{\epsilon T}\right),
\end{equation}
with $N(\epsilon)$ being the maximal number of distinct energy gaps within an interval $\epsilon$ \cite{Reimann2012}.

To characterize observable equilibration, one may define a time-dependent probability vector \(\vec{p}_t\) associated with a complete set of measurement operators \(\{P_l\}\). Each component \(p_l(t)\) represents the probability of obtaining outcome \(l\) at time \(t\) and is given by
\begin{equation}\label{eq:population_equilibration}
    p_l(t) = \tr(P_l \rho_t),
\end{equation}
where $P_l$ are measurement operators, typically projectors corresponding to a specific observable. The infinite-time average of this probability distribution is defined as $p_l(\infty) = \tr(P_l \omega).$ This quantity defines the long-term distribution of measurement outcomes and captures the steady-state behavior of the system as it reaches equilibrium.

Following \cite{meier2025emergence}, it can be shown that for on-time-average in the interval $[0,T]$, the distance between $\vec{p}_t$ and $\vec{p}_\infty$ in the $1$-norm satisfies
\begin{equation}\label{eq:meiersbound}
    \langle \| \vec{p}_t - \vec{p}_\infty \|_1 \rangle_T \leq \frac{1}{2} \sqrt{\frac{r}{d_{\text{eff}}} f(\epsilon, T)},
\end{equation}
with $r=\text{rank}(O)\leq d$ (\emph{i.e.}, the dimension of $\vec{p}_t$), and $p_l(\infty) := \lim\limits_{T \to \infty} \displaystyle\frac{1}{T} \int_0^T \tr(\rho_t P_l) \, dt.$

\subsection{Classical Statistical Complexity Measures}
Let \(\vec{p} = \{p_i\}_{i=1}^r\) be the probability vector of a random variable that takes \(r\) possible values.  The  Classical Statistical Complexity Measure of $\vec p$ is defined as \cite{lopez1995}
\begin{equation}
\mathcal{C}(\vec{p}) = H(\vec{p})D(\vec{p},\vec{\mathcal{I}}),
\label{complex.classica}   
\end{equation}
where $H(\vec{p})=-\sum_{i=1}^r p_i\log p_i$ is the observable Shannon entropy and $D\bigl(\vec p,\vec{\mathcal I}\bigr)
=\sum_{i=1}^r\Bigl(p_i-\tfrac1r\Bigr)^2$, quantifies the deviation from the uniform distribution $\vec{\mathcal I}=(1/r,\dots,1/r)$. 

Order and disorder represent two fundamental regimes in the study of physical and informational systems. The system's configuration is entirely predictable in perfectly ordered states, such as a crystal lattice, leading to minimal entropy. On the other hand, maximal disorder, exemplified by an ideal gas in thermal equilibrium, is characterized by uniform probability distributions over all accessible microstates, maximizing entropy. These extreme cases are straightforward to describe, as there is either complete structure or complete randomness. 

The classical complexity measure $\mathcal{C}(\vec{p})$ is designed to capture the richness of configurations that exist between these two extremes of order-disorder patterns. When the system is perfectly ordered, the entropy term $H(\vec{p})$ vanishes, resulting in $\mathcal{C}(\vec{p})=0$. Similarly, when the system is maximally disordered, on this scale, the disequilibrium term $D(\vec{p},\vec{\mathcal{I}})$ vanishes, again leading to $\mathcal{C}(\vec{p})=0$. Nontrivial complexity emerges only in intermediate configurations.

The Classical Statistical Complexity Measure (CSCM) is inherently dependent on both the descriptive framework adopted for a system and the scale of observation \cite{lopez1995}. Defined as a functional of a probability distribution, this measure is closely associated with the analysis of time series generated by classical dynamical systems. Its formulation is based on two essential components.
The first component is an entropy function that quantifies the informational content of the system. While the observable Shannon entropy is conventionally employed for this purpose, other generalized entropy measures may also be utilized, such as Tsallis entropy \cite{tsallis}, Escort-Tsallis \cite{escort.tsallis}, or R\'{e}nyi entropy \cite{renyi}. The second fundamental element is a distance measure defined on the space of probability distributions, designed to quantify the disequilibrium relative to a reference distribution, typically the microcanonical distribution. Various measures can serve this role, including the Euclidean distance (or, more generally, any 
p-norm \cite{bartle1995}), the Bhattacharyya distance \cite{bhatt}, and Wootters' distance \cite{wooters}. Additionally, statistical divergences such as the classical relative entropy (also known as the Kullback-Leibler divergence \cite{kl51}), the Hellinger distance \cite{matus2005hellinger}, and the Jensen-Shannon divergence \cite{lin1991divergence,Frank19} may be employed. It is worth noting that several generalized versions of complexity measures have been proposed in recent years, and these advancements have proven to be highly valuable in various areas of classical information theory \cite{Anteneodo1996,Catalan2002,Rosso2013,LopezRuiz2009,Sanudo2008,Montgomery2008,Sen2011,Sanudo2009,Moustakidis2012,SanchezMoreno2014,Calbet2001,LopezRuiz2011,Kowalski2012,Klamut20,Pokharel2018}.

The Quantum Statistical Complexity Measure (QSCM) is defined for a quantum state $\rho\in\mathcal{D}(\mathcal{H}_d)$, over an $d-$dimensional Hilbert space as the following functional of $\rho$ (see Ref. \cite{cesario2022quantum})
\begin{equation}
\mathcal{C}(\rho) =  S(\rho) D(\rho,\mathcal{I}),
\label{complexquantica}   
\end{equation}
where $S(\rho)$ is the von Neumann entropy, and $D(\rho,\mathcal{I})$ is a distinguishability (usually the trace distance) quantity between the state $\rho$ and the normalized maximally mixed state $\mathcal{I}$. Since the system evolves under closed unitary dynamics, any initial pure state remains pure at all times, and consequently, its von Neumann entropy vanishes. It then follows directly from Eq.~\eqref{complexquantica} that the quantum statistical complexity measure becomes identically zero, rendering it uninformative in this closed-system context \cite{cesario2022quantum}. Therefore, we restrict our analysis to the classical measure defined in Eq.~\eqref{complex.classica}.

\section{Observable Equilibration Complexity Measure}\label{sec:complexity_bounds}

Our primary interest lies in quantifying the degree of order and disorder relative to the equilibrium state $\omega$. Unlike typical complexity measures that use the maximally mixed state as a reference, we redefine the classical statistical complexity by considering the equilibrium state instead. This approach allows us to capture deviations not only from uniformity but also from the stabilized state that the system approaches over time.

The Observable Equilibration Complexity Measure (OECM) \( C(\vec{p}) \), as stated in Def.~\ref{def:OECM}, is designed to quantify the interplay between disorder and deviation from equilibrium (order) in the probability distribution associated with a quantum observable during its dynamical evolution. In this framework, a regime of minimal disorder corresponds to a highly localized probability vector in the eigenbasis of the observable, where the system exhibits minimal uncertainty. Consequently, the observable entropy \( H_O(\vec{p}) \) vanishes~\cite{meier2025emergence}, and the observable equilibration complexity measure \( C(\vec{p}) = H_O(\vec{p}) \Vert \vec{p} - \vec{p}_{\infty} \Vert_1 \) evaluates to zero, reflecting the absence of structural richness or dynamical tension.

In contrast, in the regime of minimal order, characterized by high uncertainty in the outcomes of the observable, the time-averaged effective probability vector \( \expval{\vec{p}}_T \) converges towards the equilibrium distribution \( \vec{p}_{\infty} \). Although the entropy \( H_O(\vec{p}) \) may be significant in this limit, the deviation from equilibrium, quantified by norms such as \( \langle\Vert \vec{p} - \vec{p}_{\infty} \Vert_1\rangle_T\) or \( \Vert \expval{\vec{p}}_T - \vec{p}_{\infty} \Vert_1 \) is small. Thus, the observable equilibration complexity measure is also small, as the system lacks any persistent dynamical structure differentiating it from thermal equilibrium.

A peak in complexity emerges in intermediate dynamical regimes, where the system is neither fully ordered nor completely equilibrated. In these states, the distribution is sufficiently delocalized to produce nonzero entropy \( H_O(\vec{p}) \), while still retaining a noticeable departure from equilibrium \( \Vert \vec{p} - \vec{p}_{\infty} \Vert_1 \gg 0 \). In such cases, \( C(\vec{p}) \) captures the transient coexistence of informational richness and nonequilibrium structure, thereby identifying regions of meaningful dynamical organization in the system's evolution.

\begin{Definition}[Observable Equilibration Complexity Measure (OECM)]\label{def:OECM}
The Observable Equilibration Complexity quantifies the extent to which the probability distribution associated with an observable deviates from its equilibrium distribution $\vec{p}_{\infty}$, and it is defined as
\begin{equation}
    C(\vec{p}) = H_O(\vec{p}) \Vert \vec{p} - \vec{p}_{\infty} \Vert_1, \label{def:classical-SC}
\end{equation}
where $H_O(\vec{p})$ is the observable entropy, given by $H_O(\vec{p}) = -\sum_{l=1}^r p_l(t)\log\left(p_l(t)\right)$, $\vec{p}$ is the probability vector associated with the observable, that is, $p_l(t) = \tr\left(P_l\rho(t)\right)$, and $\vec{p}_{\infty}$ is the infinite-time average distribution as defined in Eq.~\eqref{eq:population_equilibration}.
\end{Definition}

Within this formulation, the concept of order is thus operationally tied to the localization of $\vec{p}_t$, while disorder is associated with delocalization and convergence towards equilibrium. The measure $C(\vec{p})$ captures the dynamically relevant structures that arise in the intermediate regime between these two extremes, quantifying the degree to which the observable's distribution both exhibits uncertainty and retains memory of its initial conditions. However, it is essential to note that even when the observable exhibits substantial oscillations around the equilibrium distribution, as occurs in regular or quasi-periodic dynamics, leading to a reduced effective complexity. In such cases, despite the absence of full equilibration, the system's dynamics are less complex, as they remain confined to a limited subset of the phase space and follow predictable, structured trajectories. Conversely, for the system to effectively equilibrate, it must sufficiently explore its accessible phase space~\cite{meier2025emergence}, allowing $\vec{p}_t$ to progressively sample a broader set of configurations and thereby approach $\vec{p}_{\infty}$. The Observable Equilibration Complexity Measure thus serves as a quantitative diagnostic for tracking this equilibration process through the joint analysis of entropy production and disequilibrium decay, while also distinguishing between complex, irregular dynamics and simpler, quasi-periodic behaviors.

\begin{figure}[h!]

\centering
\subfloat[\justifying 
            Time-averaged total variation distance
            \(
                \left\langle \| \vec{p}_t - \vec{p}_\infty \|_1 \right\rangle_T.
            \)
    \label{fig:tv_mean}
]{\includegraphics[width=7.0cm]{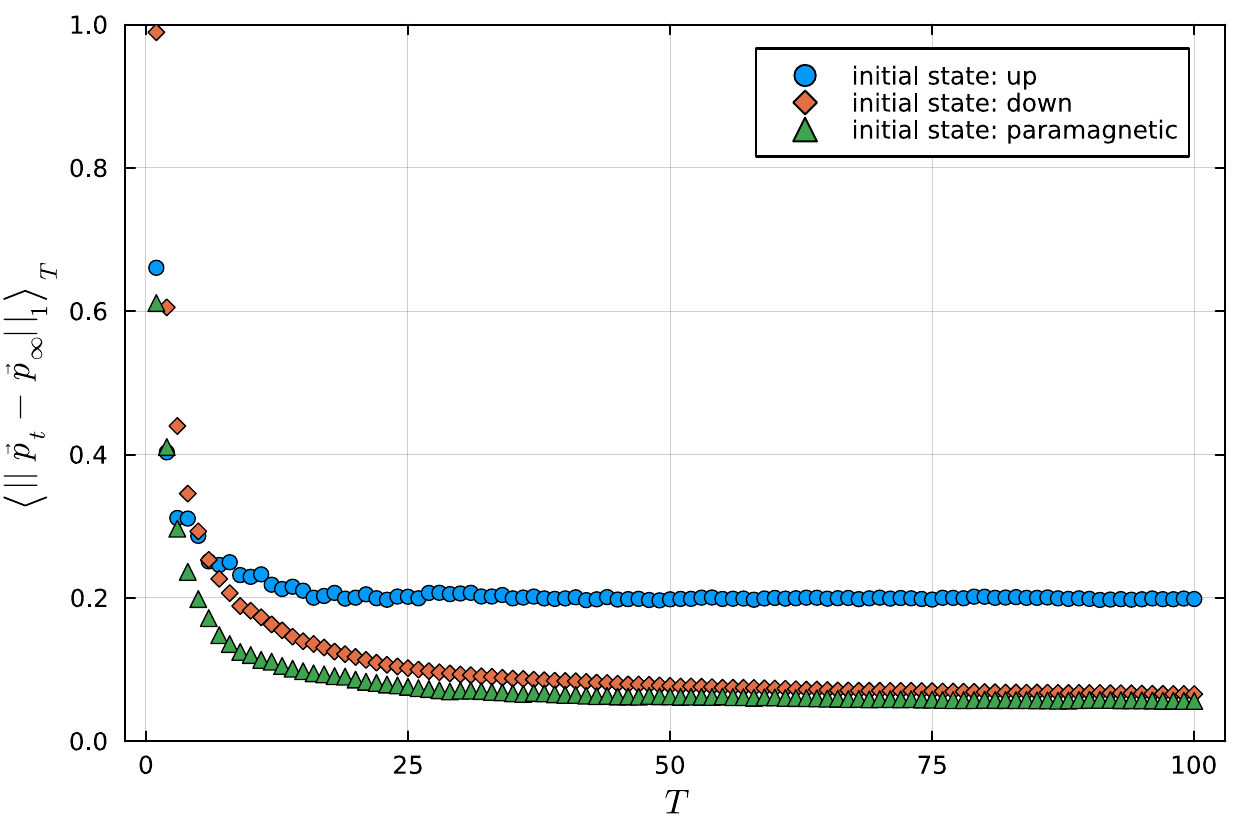}}
\hspace*{0.5cm}
\subfloat[\justifying 
            The distance between the time-averaged distribution
            \(
                \langle \vec{p}_t \rangle_T
            \)
            and the equilibrium distribution
            \(
                \vec{p}_\infty
            \).
    \label{fig:tv_avg_pt}
]{\includegraphics[width=7.0cm]{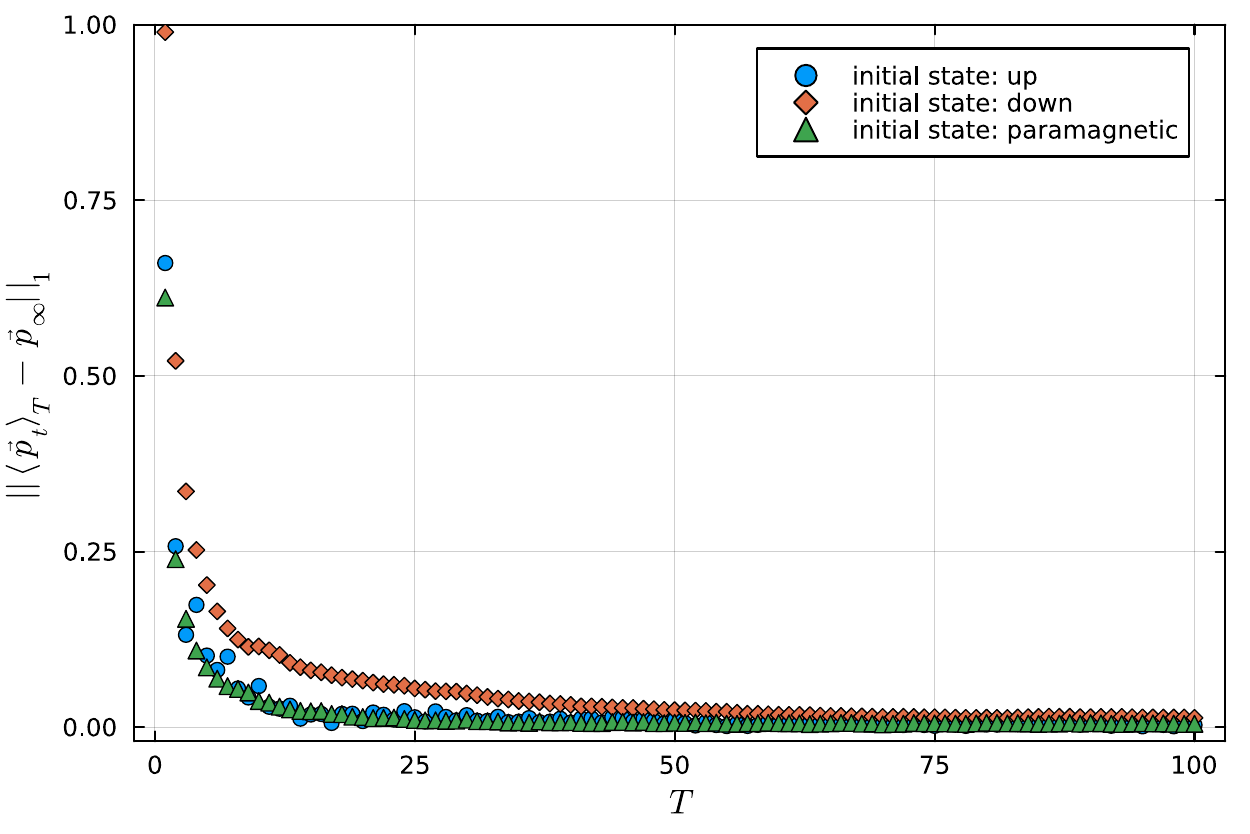}}
\caption{\justifying Comparison of convergence to equilibrium for different initial states using two distinct measures. (a) Time-averaged total variation distance $\expval{ \| \vec{p_t} - \vec{p_\infty} \|_1}_T$ for each initial state, quantifying how the instantaneous distributions approach their respective equilibrium values over time. (b) Distance between the time-averaged distribution $\expval{\vec{p}}_T$ and the equilibrium distribution $\vec{p}_\infty$, measured via the $L^1$-norm. In both panels, each initial state is represented by a distinct color: blue for the \up state, orange for the \dw state, and green for the Paramagnetic configuration, for $N=10$ spins-$1/2$.} 
\label{fig:nonconvergence}
\end{figure}

As discussed in the introduction, our goal is to define a \textit{bona fide} measure to characterize and quantify how much a given observable $O$ equilibrates under the dynamics induced by a Hamiltonian $H$. Assuming a past hypothesis where $H_O(\vec{p}(t=0)) \leq H_O(\vec{p}(t\neq 0))$, the observable equilibration complexity is expected to approach zero as the system converges towards equilibrium. Consequently, over long timescales, the time average of the complexity should tend to zero, since, on average, $\langle \| \vec{p}_t - \vec{p}_\infty \|_1 \rangle_{T\rightarrow\infty} \rightarrow 0$, as indicated by the bound in Eq.~\eqref{eq:meiersbound}. However, as illustrated in Figure~\ref{fig:tv_mean}, this is not always the case, since $\vec{p}_{t\rightarrow\infty} \neq \vec{p}_\infty$. On the other hand, in Figure~\ref{fig:tv_avg_pt}, we can observe that  $\lim_{T\rightarrow\infty} \langle \vec{p}_t \rangle_T \rightarrow \vec{p}_\infty$. This bound is a fundamental result in mathematical analysis, known as Minkowski's inequality, which extends the triangular inequality to integrals \cite{Hardy_2008}.
As one can notice, this inequality imposes a limitation on how small \( \langle \| \vec{p}_t - \vec{p}_\infty \|_1 \rangle_{T \rightarrow \infty} \) can be, and, as shown in Lemma~\ref{Lemma:variance}, this quantity is bounded from below by \( \| \langle \vec{p}_{t} \rangle_{T} - \vec{p}_{\infty} \|_1 \).

\begin{Lemma}[Variance bound on time-averaged deviation]
\label{Lemma:variance}
Let $ \rho(t) $ be a time-dependent quantum state on a finite-dimensional Hilbert space $ \mathcal{H} $, and let $\omega$ be its fixed reference state (e.g., the time-averaged state of $\rho(t)$). Let $O$ be a fixed Hermitian observable. Define the time-averaged expectation value of $ O $ over the interval $ [0,T] $ as defined in Eq.~\eqref{eq:observable_timeaverage}, the following inequality holds
\begin{equation}\label{eq:observable_variance}
\left| \langle O \rangle_T - \mathrm{Tr}[\omega O] \right|^2
\leq
\left\langle \left| \mathrm{Tr}[\rho(t) O] - \mathrm{Tr}[\omega O] \right|^2 \right\rangle_T.
\end{equation}
\end{Lemma}
\begin{proof}
Eq.~\eqref{eq:observable_variance} is a direct consequence of Minkowski's inequality. The corresponding temporal variance of a given Schur-convex function \( f(t) \) over the interval \( [0,T] \) is $\mathrm{Var}_T[f] := \left\langle |f(t)|^2 \right\rangle_T - \left| \langle f(t) \rangle_T \right|^2 \geq~0$.     
\end{proof}

To better understand how OECM behaves over time, we now derive an upper bound on its time average. This bound captures the interplay between the entropic content of measurement outcomes and their deviation from equilibrium. In particular, we show that the time-averaged complexity is controlled by the effective dimension of the initial state, the rank \(r\) of the observable, and a function \(f(\varepsilon, T)\) that encodes the system's dynamical timescales \cite{Short_2012}. Although this result does not yield a tight bound (since the observable entropy is not compared to its maximum over the constrained support), it still provides a meaningful characterization of the extent to which equilibration suppresses both uncertainty and distinguishability from the equilibrium distribution. In Appendix~\ref{app_A}, we discuss the numerical bound for this measure, stated precisely in the following theorem.

\begin{Theorem}[OECM Upper Bound]\label{theo: OECM-UB}
Let \( \vec{p}(t) \) be the probability distribution over the eigenbasis of a quantum observable \( O \) at time \( t \), and let \( \vec{p}_\infty \) be its equilibrium distribution. Define the Observable Equilibration Complexity Measure, that is, (OECM), as
\begin{equation}
C(\vec{p}(t)) = H_O(\vec{p}(t)) \Vert \vec{p}(t) - \vec{p}_\infty \Vert_1,
\end{equation}
where \( H_O(\vec{p}) \) is the observable entropy. Then, the time-averaged observable equilibration complexity \( \expval{C}_T \) satisfies the upper bound
\begin{equation}
\expval{C(\vec{p}(t))}_T \leq \frac{\log r}{2}  \sqrt{ \frac{r }{\deff }f(\varepsilon, T)},
\end{equation}
where \( r \) is the rank of the observable \( O \), \( \deff \) is the effective dimension of the initial state, and \( f(\varepsilon, T) \) is defined in ~\eqref{eq:fepsilonT}.
\end{Theorem}

\begin{proof}
By applying the Cauchy-Schwarz inequality to the time-averaged observable equilibration complexity 
\begin{equation}
\expval{C(\vec{p}(t))}_T = \expval{ H_O(\vec{p}(t)) \Vert \vec{p}(t) - \vec{p}_\infty \Vert_1 }_T \leq \sqrt{ \expval{ H_O^2(\vec{p}(t)) }_T  \expval{ \Vert \vec{p}(t) - \vec{p}_\infty \Vert_1^2 }_T }.
\end{equation}
Since the observable entropy \( H_O(\vec{p}) \leq \log r \), we have
\begin{equation}
\expval{ H_O^2(\vec{p}(t)) }_T \leq (\log r)^2.
\end{equation}
From the equilibration bound derived in~Appendix 1 of \cite{meier2025emergence}, one obtains
\begin{equation}
\expval{ \Vert \vec{p}(t) - \vec{p}_\infty \Vert_1^2 }_T \leq \frac{r }{4\deff}f(\varepsilon, T).
\end{equation}
Substituting both results into the previous inequality, we find
\begin{equation}
\label{eq:bound_C}
\expval{C(\vec{p}(t))}_T \leq  \frac{\log r}{2}  \sqrt{ \frac{ r }{\deff}f(\varepsilon, T)}.
\end{equation}
This completes the proof.
\end{proof}

This approach results in a bound that is not particularly restrictive, since the rank of the observable \(r\) does not impose a strong limitation on the equilibration complexity. Therefore, while the bound is useful for estimating the system's behavior, it does not provide a precise and tight description of the equilibration process, as the entropy associated with the observable is not the maximum possible for all states. The bound described in Eq.~\eqref{eq:bound_C} holds for large times $T$, specifically for $T \gg \frac{8 \log_2(n)}{\varepsilon}$, where the function $f(\epsilon, T)$, that captures properties of the energy spectrum
$f(\epsilon, T) = N(\epsilon) \left(1 + \frac{8 \log_2(n)}{\epsilon T}\right)\approx 1$, 
with \( N(\varepsilon) \) being the maximal number of distinct energy gaps within an interval \( \varepsilon \)~\cite{Reimann2012}. As discussed in Ref.~\cite{Short_2012}, in this asymptotic regime, the bound simplifies to $\left\langle C(\vec{p}_t) \right\rangle_T \lesssim \frac{\log r}{2} \sqrt{ \frac{r}{d_{\mathrm{eff}}} },$ revealing a direct inverse square-root dependence on the effective dimension. This behavior underscores the role of quantum state delocalization in suppressing observable complexity at equilibrium. A detailed numerical analysis of this behavior is presented in Appendix~\ref{app_A}.

In the study of quantum equilibration, it is often assumed that the time-averaged deviation \( \langle \|\vec{p}_t - \vec{p}_\infty\|_1 \rangle_T \) vanishes as time increases, indicating convergence to equilibrium. However, this assumption generally holds only in the thermodynamic limit \( N \to \infty \). In finite systems, especially those exhibiting slow or incomplete equilibration, the instantaneous probability distribution \( \vec{p}_t \) may not converge to the equilibrium distribution \( \vec{p}_\infty \), as shown in Figure~\ref{fig:tv_mean}. Nevertheless, the quantity \( \|\langle \vec{p}_t \rangle_T - \vec{p}_\infty \|_1 \) tends to zero and offers a robust alternative for characterizing equilibration. Building on this, we introduce the Time-Averaged Observable Equilibration Complexity Measure,  $C(\langle  \vec{p}_{t}\rangle_{T})$ which quantifies how far the system remains from equilibrium by combining the observable entropy \( H_O(\langle \vec{p}_t \rangle_T) \) with the trace distance \( \|\langle \vec{p}_t \rangle_T - \vec{p}_\infty\|_1 \), as seen in Figure~\ref{fig:tv_avg_pt}. This measure vanishes when the system is either in a pure measurement outcome or has equilibrated on average, and it captures the interplay between coherence, uncertainty, and relaxation dynamics in a compact, physically meaningful way.

In Definition~\ref{def: timeaverage_complexity}, we present the \emph{Time-Average Observable Equilibration Complexity Measure}, which quantifies this evolution by combining the observable Shannon entropy of the time-averaged probability vector and the distance to the equilibrium distribution. This measure captures the complexity of the equilibration process, being zero when the system is in a pure state or has fully equilibrated.

\begin{Definition}[Time-Average Observable Equilibration Complexity Measure]\label{def: timeaverage_complexity}
Considering, in a time interval $[0,T]$, the time average probability vector $\langle  \vec{p}_{t}\rangle_{T}$ with elements $\langle p_l(t)\rangle_{T} = \langle\tr(P_l \rho_t)\rangle_{T}$, such that $\sum_l \langle p_l(t)\rangle_{T} = 1$. We can define the \emph{Time-Average Observable Equilibration Complexity Measure} as
\begin{equation}\label{eq:timeaverage_complexity}
     C(\langle  \vec{p}_{t}\rangle_{T})  = H(\langle  \vec{p}_{t}\rangle_{T})\Vert \langle  \vec{p}_{t}\rangle_{T} - \Vec{p}_{\infty} \Vert_1,
\end{equation}
which is zero for $\vec{p}_{t}$ that are pure probability vectors or when they approach equilibrium $\lim_{T\rightarrow\infty}\langle  \vec{p}_{t}\rangle_{T} = \Vec{p}_{\infty}$.
\end{Definition}
In Appendix~\ref{app_A}, we discuss the numerical bound for this measure; however, Definition~\ref{def: timeaverage_complexity} can be viewed as a particular instance of Definition~\ref{def:OECM}. Within this framework, it satisfies the upper bound presented in Eq.~\eqref{eq:bound_C} of Theorem~\ref{theo: OECM-UB}. 

The Time-Average Observable Equilibration Complexity Measure converges as  $\Vert \langle  \vec{p}_{t}\rangle_{T} -~ \Vec{p}_{\infty} \Vert_1$ converges to zero. Now, we compute a saturation bound for the probability of $\vec{p}_t$ approaching $\langle  \vec{p}_{t}\rangle_{T}$ in the limit of $T\rightarrow \infty$. Theorem~\ref{theorem:equilibrium_deviation} provides a probabilistic bound on the deviation between the time-averaged measurement statistics and their equilibrium values. It shows that large deviations are unlikely when the effective dimension is high.

\begin{Theorem}[Equilibrium Deviation Bound]\label{theorem:equilibrium_deviation}
Consider a random initial state \( \rho_0 \) drawn from an ensemble with effective dimension \( \deff \), evolving under a Hamiltonian with non-degenerate energy gaps. For any \begin{equation}
\mathbb{P}_{\rho_0} \left( \left\| \langle \vec{p}_t \rangle_T - \vec{p}_\infty \right\|_2 \geq  \varepsilon \right) \leq \frac{1}{\varepsilon^2}\frac{r}{\deff \, }f(\epsilon,T).
\end{equation}
where \( r \) is the number of measurement outcomes.
\end{Theorem}
\begin{proof}
For each component \( p_l(t) \), Jensen's inequality yields
\begin{equation}
\left( \langle p_l(t) \rangle_T - p_l^\infty \right)^2 \leq \frac{1}{T} \int_0^T \left( p_l(t) - p_l^\infty \right)^2 dt.
\end{equation}
Taking the expectation over pure states \( \rho_0 = \ketbra{\psi(0)}{\psi(0)} \), where \( \ket{\psi(0)} \) is sampled according to the Haar measure on the Hilbert space, and applying Fubini's theorem, we obtain
\begin{equation}
\mathbb{E}_{\rho_0} \left[ \left( \langle p_l(t) \rangle_T - p_l^\infty \right)^2 \right] 
\leq  \mathbb{E}_{\rho_0} \left[\langle\left( p_l(t) - p_l^\infty \right)^2\rangle_T \right]. \end{equation}
On the other hand, applying Riemann's bound, for $p_l(t)=\tr(P_l\rho(t))$ with $||P_l||=1$, we have
\begin{equation}
 \left\langle \left( p_l(t) - p_l^\infty \right)^2 \right\rangle_T \leq \frac{f(\epsilon,T)}{\deff}.
\end{equation}
Therefore,
\begin{equation}
  \mathbb{E}_{\rho_0} \left[ \left( \langle p_l(t) \rangle_T - p_l^\infty \right)^2 \right]   \leq  \frac{f(\epsilon,T)}{\deff},  
\end{equation}
as $\mathbb{E}_{\rho_0} \left[\frac{f(\epsilon,T)}{\deff}\right] =\frac{f(\epsilon,T)}{\deff}.$ Summing the elements, we obtain the $l_2$ distance 
\begin{equation}
    \mathbb{E}_{\rho_0} \left[ \| \langle \vec{p}_{t} \rangle_T - \vec{p}_{\infty} \|_2^2 \right] =
    \sum_l \mathbb{E}_{\rho_0} \left[ \left( \langle p_l(t) \rangle_T - p_l^\infty \right)^2 \right]  \leq  \frac{r}{\deff}f(\epsilon,T).
\end{equation}
Applying Markov's inequality,
\begin{equation}
\mathbb{P}_{\rho_0} \left( \left\| \langle \vec{p}_t \rangle_T - \vec{p}_\infty \right\|_2 \geq \varepsilon \right)
\leq \frac{1}{\varepsilon^2} \mathbb{E}_{\rho_0} \left[ \left\| \langle \vec{p}_t \rangle_T - \vec{p}_\infty \right\|_2^2 \right]
\leq \frac{r}{\deff \, \varepsilon^2}f(\epsilon,T),
\end{equation}
and this completes the proof.
\end{proof}

\section{Numerical Applications}\label{sec:numerics}

The Hamiltonian governing the time evolution of the system is a spin-$\frac{1}{2}$ Ising-like model incorporating both longitudinal and transverse magnetic fields, expressed as
\begin{equation}
H = g \sum_{i=1}^{N} \hat{\sigma}_i^x + h \sum_{i=2}^{N-1} \hat{\sigma}_i^z + J \sum_{i=1}^{N-1} \hat{\sigma}_i^z \hat{\sigma}_{i+1}^z + (h - J) \left( \hat{\sigma}_1^z + \hat{\sigma}_N^z \right),
\end{equation}
where $\hat{\sigma}_i^\alpha$ with $\alpha = x, y, z$ denote the Pauli spin operators acting on site $i$ of the chain. The parameters $g$ and $h$ correspond to the strengths of the transverse and longitudinal magnetic fields, respectively, while $J$ defines the strength of the spin-spin interaction coupling \cite{Banuls11}.

In the simulations, the values of the model parameters were selected to emphasize the \textit{non-integrable} regime, specifically $g = \frac{5 + \sqrt{5}}{8}, \quad h = \frac{1 + \sqrt{5}}{4}, \quad \text{and} \quad J = 1$, see Ref. \cite{Yamaguchi24}. 
These parameter choices are consistent with those employed in previous studies on equilibration and thermalization in isolated quantum systems, thereby ensuring a rich and nontrivial dynamical behavior~\cite{meier2025emergence,reimann2018dynamical}. For the numerical analysis, we considered the following initial states: the fully polarized \up state, $\ket{\uparrow \uparrow \dots \uparrow}$ (\texttt{Up}); the fully polarized \dw state, $\ket{\downarrow \downarrow \dots \downarrow}$ (\texttt{Dw}); and the paramagnetic configuration, $\ket{\uparrow \downarrow \uparrow \downarrow \dots}$ (\texttt{Pm}), for a chain of $N = 10$ spins-$1/2$ particles.

Given an initial state composed of $N$ spins-$\frac{1}{2}$ particles and the Hamiltonian $H$ defined above, we perform the unitary time evolution according to the Schr\"{o}dinger equation using the \texttt{QuantumOptics.jl} library in Julia. The time-dependent state of the system is, thus, obtained as
\begin{equation}
\ket{\psi(t)} = e^{-i H t} \ket{\psi(0)}.
\end{equation}

Subsequently, we compute the equilibrium state $\omega$ via exact diagonalization, corresponding to the infinite-time average of the evolved state. For a specific observable, namely the magnetization \textit{per} particle, we monitor its time evolution to analyze relaxation and equilibration phenomena.
\begin{equation}
M_z(t) = \frac{1}{N} \sum_{i=1}^{N} \langle \sigma_z^{(i)} \rangle (t).
\end{equation}

\begin{figure}[h!]
\centering

\subfloat[\justifying
    Time evolution of the magnetization \(M_z(t)\).
    \label{fig:magnetization_time}
]{\includegraphics[width=7.0cm]{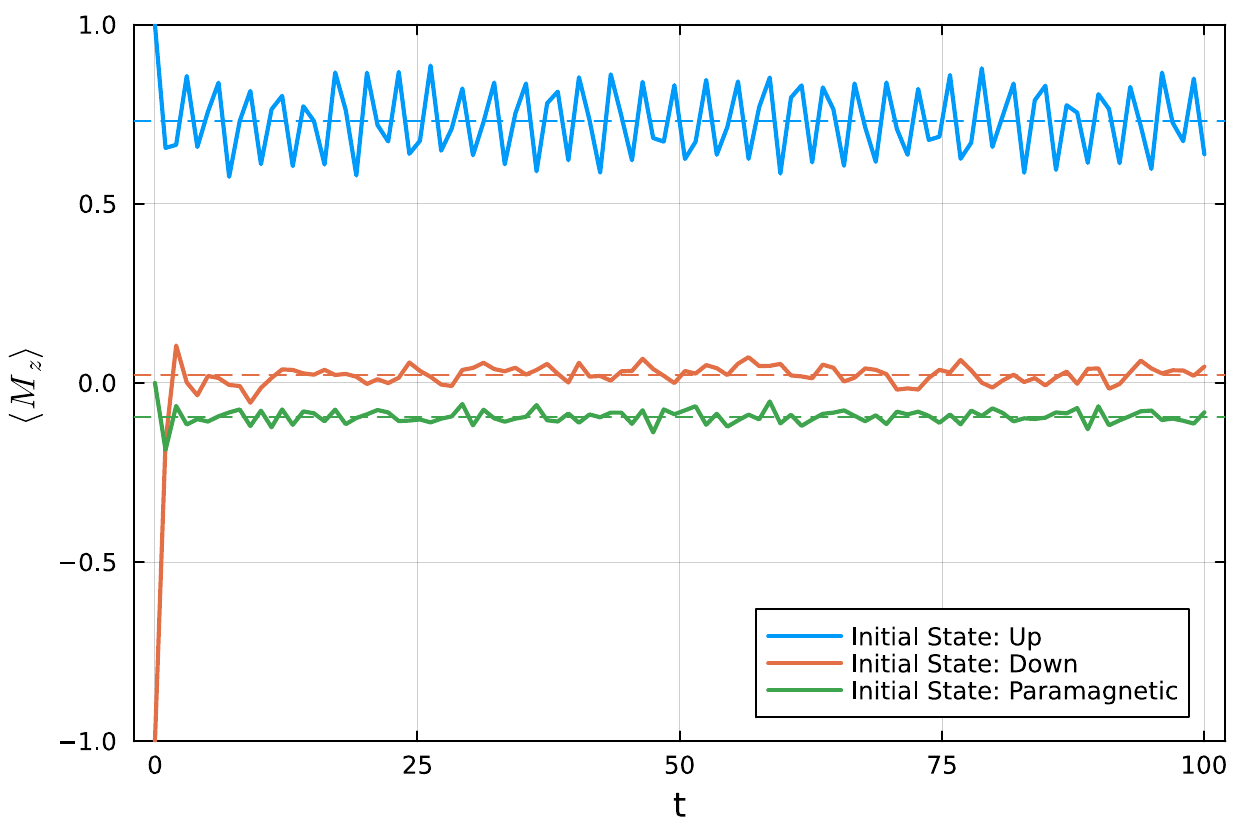}}
\hspace*{0.5cm}
\subfloat[\justifying
    Time-averaged magnetization \(\langle M_z \rangle_T\).
    \label{fig:magnetization_avg}
]{\includegraphics[width=7.0cm]{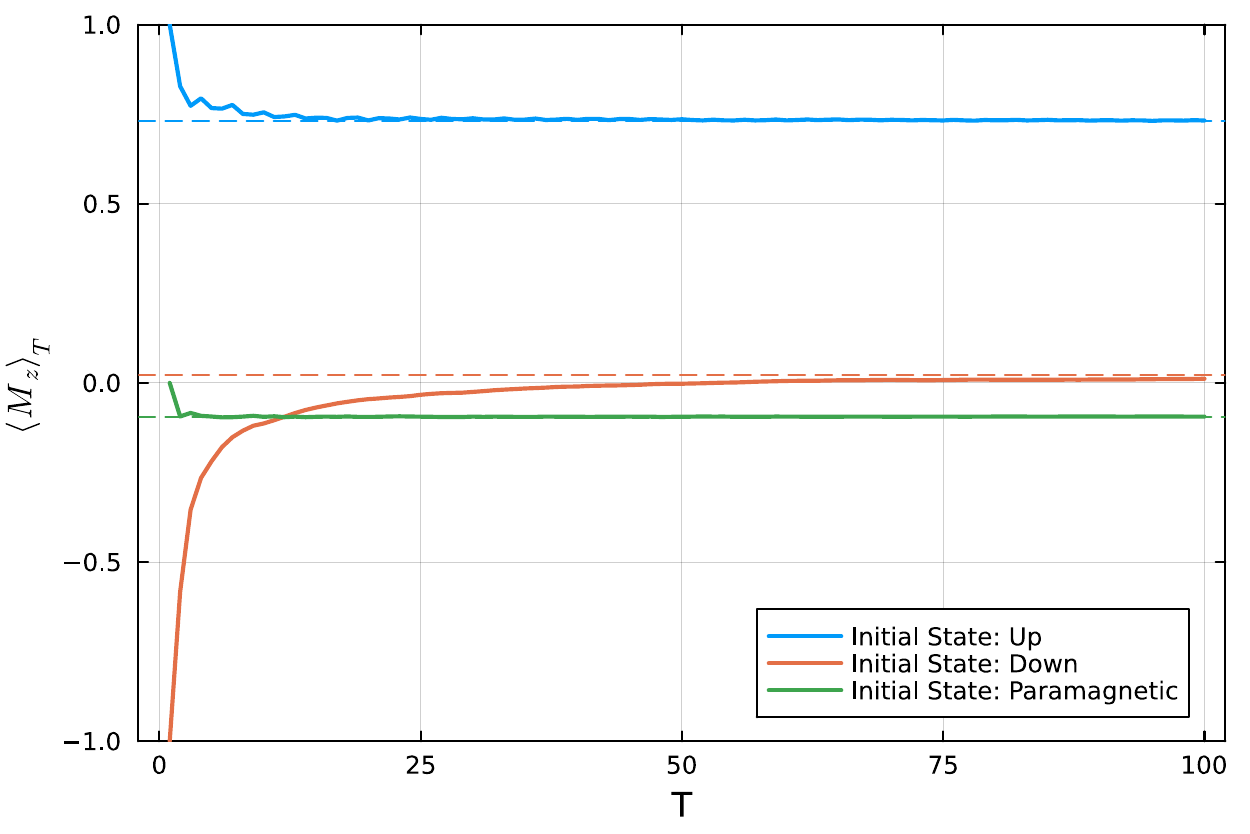}}

\caption{\justifying 
Magnetization dynamics for different initial states. (\textbf{a}) Instantaneous evolution of \(M_z(t)\) over time. (\textbf{b}) Convergence of the time-averaged magnetization \(\langle M_z \rangle_T\) to its corresponding equilibrium value as a function of \(T\). In both panels, each initial state is represented by a distinct color: blue for \texttt{Up}, orange for \texttt{Dw}, and green for the \mypm configuration. The dashed lines in matching colors indicate the equilibrium values associated with each initial state, for \(N=10\) spins-\(\frac{1}{2}\).}
\label{fig:magnetization_dynamics}
\end{figure}

Figure~\ref{fig:magnetization_dynamics} presents the temporal evolution of the magnetization \textit{per particle} $M_z(t)$ for different initial states: the fully polarized up state (\texttt{Up}), the fully polarized down state (\texttt{Dw}), and an alternating paramagnetic configuration (\texttt{Pm}). In Figure~\ref{fig:magnetization_time}, we observe that each initial condition evolves distinctly, exhibiting characteristic oscillations before tending towards stabilization around a mean value. Figure~\ref{fig:magnetization_avg} shows the convergence of the time-averaged magnetization $\langle M_z \rangle_T$ towards its corresponding equilibrium value, indicated by dashed lines. These results confirm the occurrence of an equilibration process, whereby the system, despite being closed and evolving unitarily, displays relaxation of observables towards stable values.

\begin{figure}[h!]
\centering

\subfloat[\justifying
    Time evolution of the observable Shannon entropy \(H_O(t)\).
    \label{fig:entropy_time}
]{\includegraphics[width=7.0cm]{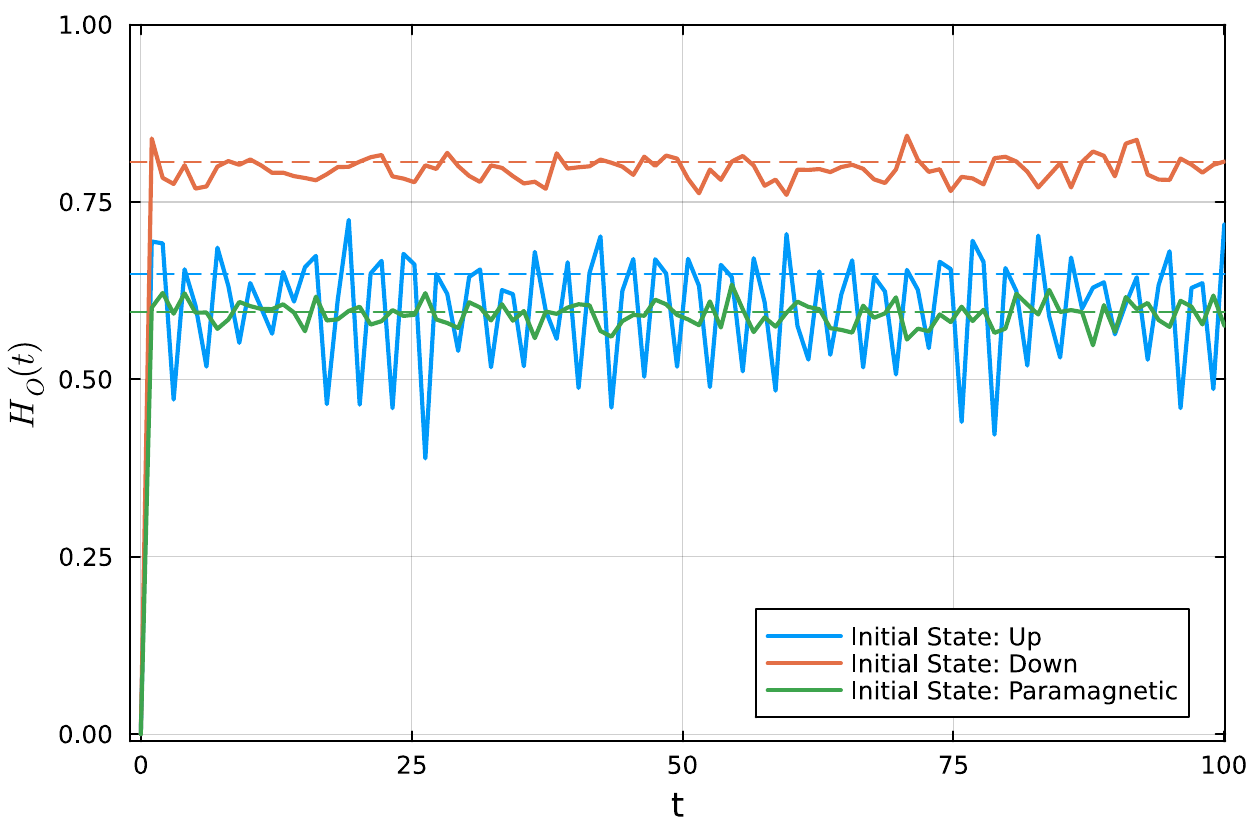}}
\hspace*{0.5cm}
\subfloat[\justifying
    Convergence of the time-averaged observable Shannon entropy \(\langle H_O \rangle_T\).
    \label{fig:entropy_avg}
]{\includegraphics[width=7.0cm]{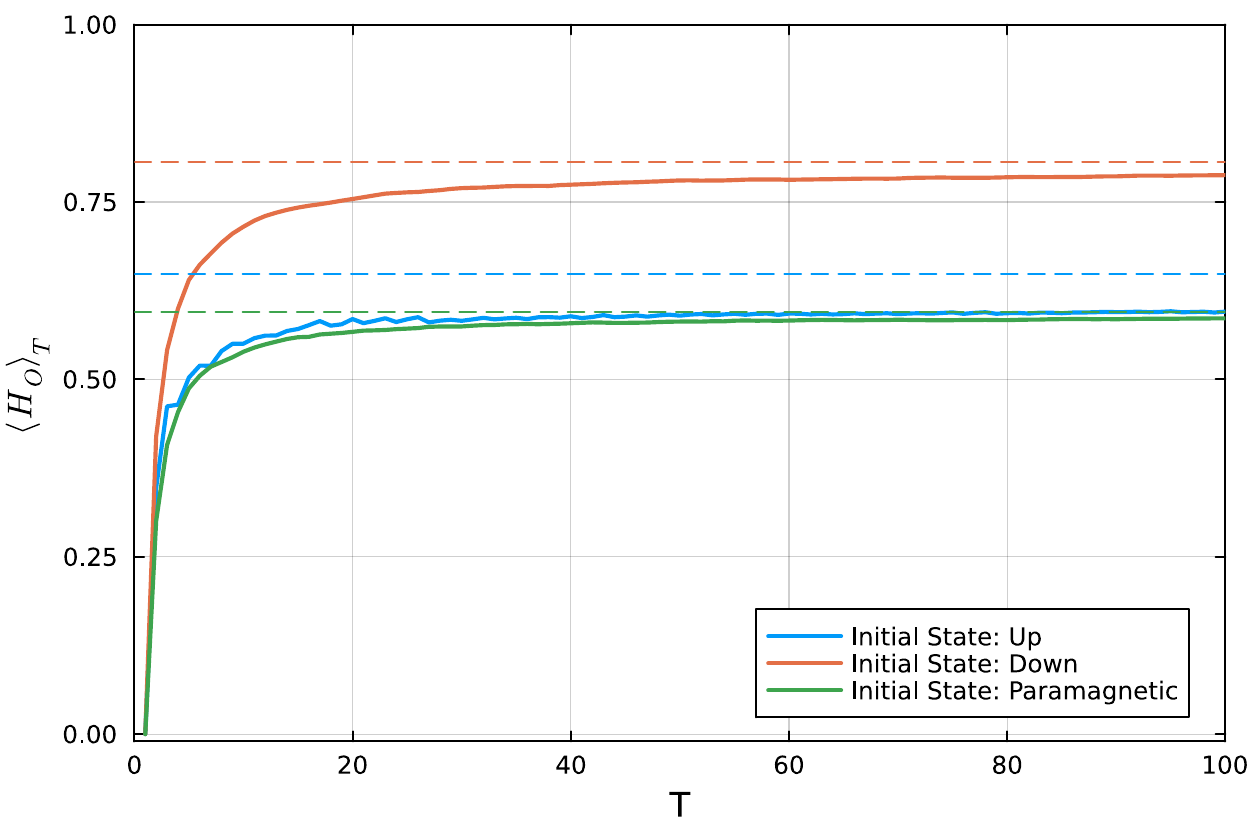}}

\caption{\justifying 
Evolution of the observable entropy for different initial states. (\textbf{a}) Instantaneous behavior of the observable Shannon entropy \(H_O(p_t)\) associated with the probability distribution of the measurement outcomes. (\textbf{b}) Convergence of the time-averaged observable entropy \(\langle H_O \rangle_T\) as a function of \(T\). In both panels, each initial state is represented by a distinct color: blue for the \up state, orange for the \dw state, and green for the Paramagnetic configuration. Dashed lines in matching colors indicate the equilibrium entropy values obtained from the stationary distribution. Entropy values are reported with a logarithmic base selected to normalize the entropy of the maximally mixed distribution to unity, assuming a system of \( N = 10 \) spin-\(\frac{1}{2} \) particles.
}\label{fig:entropy_dynamics}
\end{figure}

\up state exhibits a more regular and quasi-periodic behavior, as depicted in Figure~\ref{fig:magnetization_dynamics}. This distinctive dynamical pattern can be attributed to its relatively low effective dimension ($d_{\text{eff}}^{\up} \approx 2.95$), which severely restricts the extent to which the state can explore the available Hilbert space. In comparison, the Down and Paramagnetic states possess significantly higher effective dimensions, with $d_{\text{eff}}^{\dw}\approx 93.74$ and $d_{\text{eff}}^{\mypm} \approx 23.25$, respectively. These larger effective dimensions facilitate more extensive mixing among energy eigenstates, thereby promoting richer and more complex dynamics, as will be further elucidated in Figure~\ref{fig:entropy_dynamics}. Consequently, the Down and Paramagnetic configurations exhibit dynamical behaviors that are characteristic of equilibration, with the system's observables progressively relaxing towards their equilibrium values.

Through the procedure described above, we can monitor the time evolution of the magnetization and track the full probability distribution of measurement outcomes at each instant. This allows for the computation of the observable Shannon entropy associated with the observable, often referred to as the observable entropy, which quantifies the degree of uncertainty or disorder in the system at a given time. 

In Figure~\ref{fig:entropy_dynamics}, we analyze the time evolution of the observable entropy \( H_O(t) \), which quantifies the uncertainty in the probability distributions associated with the magnetization measurement. In panel~(a), all initial states begin with zero entropy, reflecting complete predictability in the measurement basis. As time progresses, the entropy increases due to the spreading of the probability distribution, signaling a loss of information about the measurement outcomes and a tendency toward equilibrium.

Panel~(b), however, reveals that the time-averaged observable entropy \( \langle H_O \rangle_T \) for the \up state does not converge to its equilibrium value \( H_O^\infty \). This indicates that the probability distribution remains partially localized even after long-time evolution. The origin of this persistent deviation lies in the low effective dimension of  \up state, which restricts its dynamics to a narrow subset of the Hamiltonian's eigenbasis. As a result, the evolution remains highly structured and quasi-periodic, preserving a significant amount of informational order over time.

It is noteworthy that although the \up and \mypm states may exhibit similar instantaneous entropy in the observable basis, their dynamical behaviors differ fundamentally. The effective dimension of the \mypm state is substantially higher, allowing for broader spreading across the energy spectrum and enabling stronger equilibration of observable statistics. This highlights that observable entropy alone may be insufficient to characterize equilibration; the effective dimension and the spectral structure of the initial state also play a critical role. These quantities also exhibit features consistent with the framework of emergent equilibration laws based on constrained subspace dynamics, as discussed in Ref.~\cite{meier2025emergence}, where the long-time behavior of population observables reflects a form of the second law of thermodynamics in closed quantum systems.

\begin{figure}[h!]
\centering

\subfloat[\justifying
    Time-averaged of the OECM \(\langle C(\vec{p_t}) \rangle_T\).
    \label{fig:mean_C_t}
]{\includegraphics[width=7.0cm]{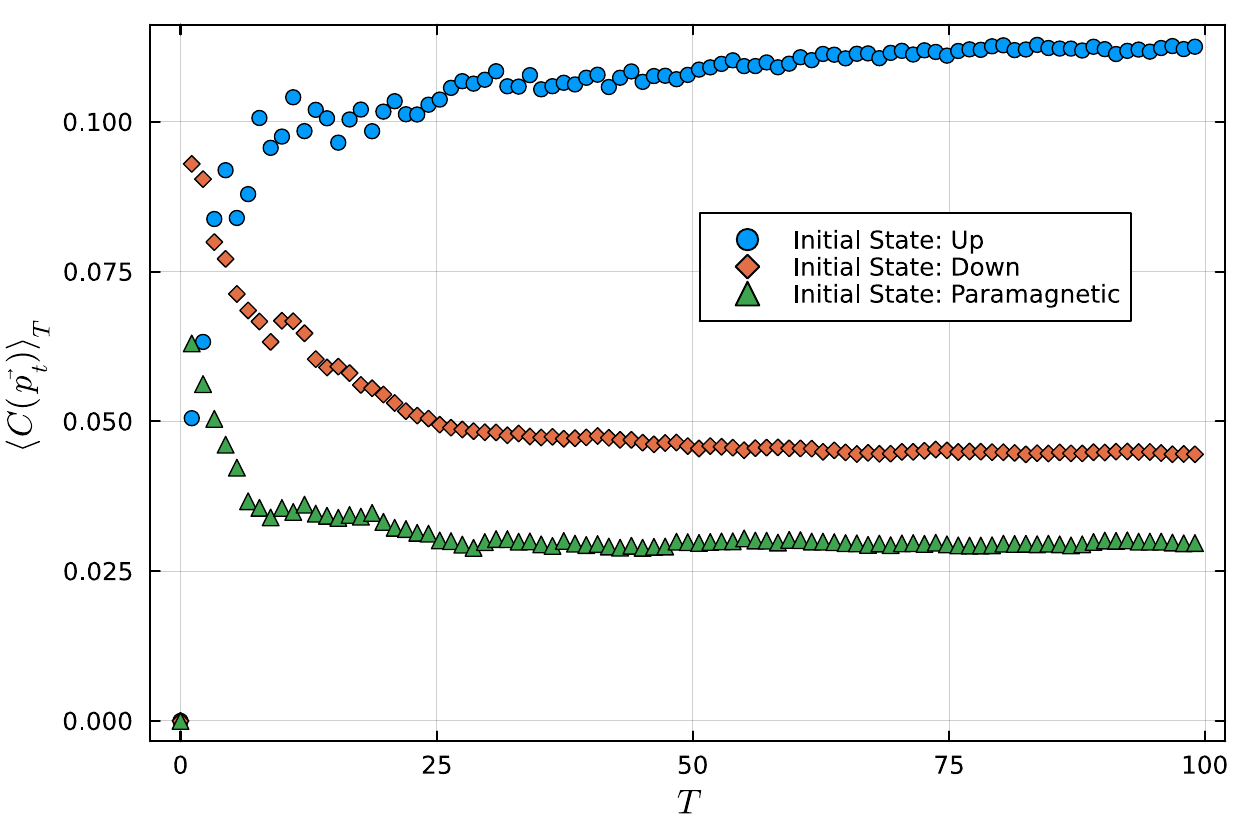}}
\hspace*{0.5cm}
\subfloat[\justifying
    OECM of the time-averaged distribution \(C(\expval{\vec{p_t}}_T)\).
    \label{fig:C_of_mean}
]{\includegraphics[width=7.0cm]{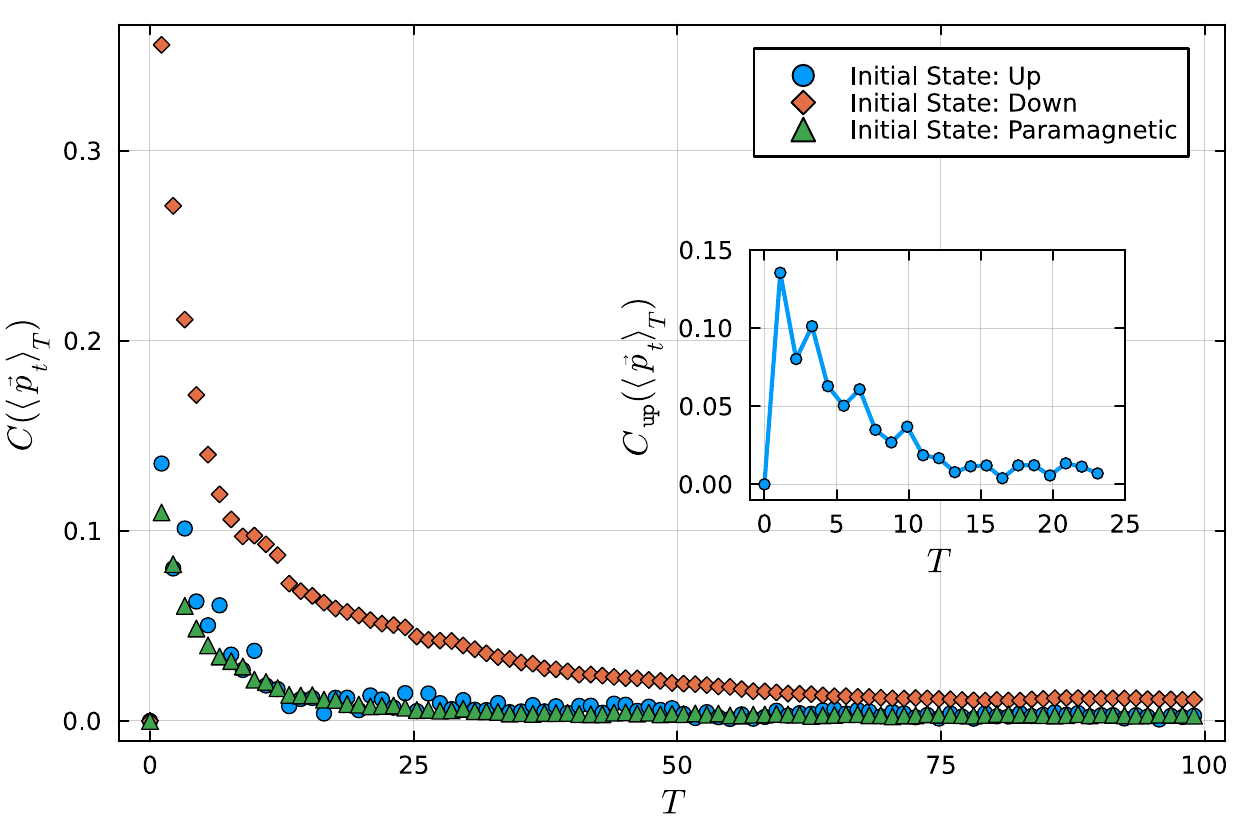}}

\caption{\justifying 
Comparison between the time-averaged observable equilibration complexity and the observable equilibration complexity of the time-averaged distribution, each as a function of time \(T\). (\textbf{a}) \(\langle C(\vec{p_t}) \rangle_T\), quantifying the average over time of the product between the observable entropy and the trace distance to equilibrium. (\textbf{b}) Observable equilibration complexity evaluated at the time-averaged distribution, \(C(\expval{\vec{p_t}}_T)\). Each curve corresponds to a distinct initial state: blue for the \up state, orange for \texttt{Dw}, and green for \mypm (Paramagnetic configuration), for \(N=10\) spins-\(\frac{1}{2}\). A magnified inset in panel \textbf{(b)} focuses on the \up initial state, revealing subtle, but enduring fluctuations in \(C(\langle \vec{p}_t \rangle_T)\).}
\label{fig:comparison_time_average}
\end{figure}

Figure~\ref{fig:mean_C_t} illustrates the average observable complexity \(\langle C(\vec{p}_t) \rangle_T\), which quantifies the typical instantaneous deviation of the observable distribution from equilibrium. This measure captures how dynamically active and structured the observable remains throughout the evolution. The \up state exhibits the highest average observable equilibration complexity, remaining persistently, \textit{i.e.} pointwise, far from equilibrium with significant fluctuations in the observable distribution. This reflects a regime of strong, coherent revivals and poor dephasing, characteristic of a low effective dimension.

In contrast, the \mypm and \dw states exhibit a rapid suppression of \(\langle C(\vec{p}_t) \rangle_T\), suggesting a faster approach to equilibrium-like behavior in the instantaneous statistics. However, while the \mypm state exhibits both low entropy and relatively small deviations from equilibrium, the \dw state remains more complex due to its persistently high entropy. As shown in Figure~\ref{fig:entropy_dynamics}, the \dw state's entropy increases rapidly. Still, it stabilizes at a value significantly higher than that of the Paramagnetic configuration, indicating a broader and more uncertain observable distribution. At the same time, Figure~\ref{fig:nonconvergence} reveals that neither the \dw nor the \mypm state fully equilibrates pointwise, as the instantaneous distribution \(\vec{p}_t\) retains a finite distance from \(\vec{p}_\infty\) even at long times. Consequently, \(\langle C(\vec{p}_t) \rangle_T\) remains nonzero in the asymptotic regime for both states.

This behavior highlights that the observable equilibration complexity measure captures two complementary sources of deviation from equilibrium: informational disorder, quantified by entropy, and temporal fluctuations, quantified by the lack of pointwise convergence. In the case of the \dw state, it is the combination of high entropy and residual dynamical structure that sustains a larger value of \(\langle C(\vec{p}_t) \rangle_T\). For the \mypm state, despite its lower entropy, the persistence of small but non-vanishing fluctuations prevents the complexity from vanishing entirely, reflecting incomplete equilibration in the instantaneous statistics.

Figure~\ref{fig:C_of_mean} reveals another important distinction in the observable equilibration behavior of the initial states. In particular, we examine the Observable Equilibration Complexity Measure (OECM) of the time-averaged distribution, \( C(\langle \vec{p}_t \rangle_T) \). This measure provides a complementary perspective on the equilibration process, allowing us to quantify persistent informational structures encoded in the effective probability distribution. Although \dw state possesses the highest effective dimension among all initial configurations, it also exhibits the largest observable complexity \( C(\langle \vec{p}_t \rangle_T) \) at finite times. This apparent paradox is reconciled by noting that a high effective dimension guarantees equilibration in the long-time limit, but not necessarily a rapid decay of observable structure. The \dw state explores a broad energy subspace, leading to slow dephasing and a time-averaged observable distribution that retains nontrivial structure. As a result, the quantity \( C(\langle \vec{p}_t \rangle_T) \), which quantifies both the entropy and distinguishability of this distribution, remains elevated over the observed time window.

In contrast, the \up state, despite its low effective dimension and failure to equilibrate, exhibits a low value of \( C(\langle \vec{p}_t \rangle_T) \). This is because its evolution remains confined to a small invariant subspace, generating a time-averaged distribution that is highly structured but dynamically simple. Importantly, this low complexity does not reflect proximity to equilibrium; rather, it results from persistent quasi-periodic behavior and strong memory of the initial state. These coherent revivals lead to a time-averaged distribution that lacks both entropy and mixing, hence producing a small value of \( C(\langle \vec{p}_t \rangle_T) \), despite the absence of equilibration.

The findings underscore that a low value of observable complexity in the time-averaged sense does not necessarily indicate relaxation, and that a high effective dimension, while necessary for equilibration, can delay the suppression of structure in the observable distribution. The observable equilibration complexity measure \( C(\langle \vec{p}_t \rangle_T) \) thus serves as a sensitive witness of equilibration structure and temporal memory. Taken together with Figure~\ref{fig:mean_C_t}, these results reveal a fundamental distinction between dynamical persistence and structural convergence in quantum equilibration. The time-averaged complexity \(\langle C(\vec{p}_t) \rangle_T\) captures the degree to which the system fluctuates away from equilibrium at each instant in time, whereas the complexity of the time-averaged distribution, \(C(\langle \vec{p}_t \rangle_T)\), reflects the cumulative structure retained by the observable statistics over time.

For example, while the \dw state exhibits high values of both measures (due to its slow dephasing and broad exploration of the Hilbert space), the \up state displays an intriguing asymmetry: it is dynamically active, as shown by its large \(\langle C(\vec{p}_t) \rangle_T\), yet structurally simple in the time-averaged sense, with a small and quasi-periodic \(C(\langle \vec{p}_t \rangle_T)\). This contrast arises because its evolution is confined to a small invariant subspace, leading to persistent quasi-periodic behavior and low entropy accumulation.

These observations emphasize the complementary roles of \(\langle C(\vec{p}_t) \rangle_T\) and \(C(\langle \vec{p}_t \rangle_T)\): the former probes the persistence of temporal non-equilibrium, while the latter detects structural memory in the observable distribution. Their combined analysis provides a refined diagnosis of equilibration behavior that goes beyond traditional measures based solely on effective dimension or asymptotic values.

\section{Conclusions}\label{sec:conclusion}

In this work, we investigated the role of statistical complexity as a diagnostic tool for analyzing observable equilibration in isolated quantum systems evolving under unitary dynamics. To this end, we introduced the Observable Equilibration Complexity Measure, defined as the product of the observable entropy and the trace distance to equilibrium. This construction provides a formal framework for quantifying the transient informational structures that arise during the system's approach to equilibrium.

Our theoretical developments established upper bounds on the time-averaged complexity in terms of the effective dimension and spectral properties of the system. These predictions were corroborated by numerical simulations of a non-integrable Ising-like spin chain Hamiltonian. For initial states with high effective dimension, such as the \dw and \mypm configurations, the system exhibited clear signatures of equilibration, with the complexity vanishing as the time-averaged probability vector converged to the equilibrium distribution. In contrast, for initial states with low effective dimension, such as the \up state, we observed a non-null  \( \langle C(\vec{p}_t) \rangle_T \)  due to the non-pointwise convergence of the probability vector. 

Furthermore, we evaluated the complexity measure of the effective distribution \( \langle \vec{p}_t \rangle_T \) through \( C(\langle \vec{p}_t \rangle_T) \), which converges after a certain time threshold. We observed the persistence of a quasi-periodic regime for the \up state, which is consistent with the interpretation that a state with lower effective dimension has less difficulty in retaining information about the initial state and, as a result, explores only a limited portion of the Hamiltonian subspaces, leading to revivals in the complexity of the effective distribution.

Moreover, the analysis of the observable complexity, both in its instantaneous average form \( \langle C(\vec{p}_t) \rangle_T \) and in the complexity of the time-averaged distribution \( C(\langle \vec{p}_t \rangle_T) \), captures distinct and complementary aspects of quantum equilibration. While a highly effective dimension is generally associated with equilibration, it does not guarantee rapid suppression of observable structure. The \dw state exemplifies this by exhibiting large complexity due to the coexistence of high entropy and persistent dynamical features. Conversely, the \up state, despite failing to equilibrate, shows low time-averaged complexity as a result of its confinement to a small invariant subspace with coherent quasi-periodic behavior. These findings emphasize that the proposed complexity measures offer refined probes of both informational disorder and temporal memory in isolated quantum systems.

In summary, our results demonstrated that the observable entropy serves as a reliable indicator of the transition from ordered to disordered regimes in the measurement statistics. At the same time, the complexity measure effectively captures the interplay between this increasing disorder and the relaxation towards equilibrium. Overall, this study advances the understanding of how classical-like equilibrium behavior emerges from unitary quantum dynamics, positioning our measure as a quantitative witness of equilibration and memory retention in closed quantum systems. Our findings suggest potential applications in characterizing equilibration phenomena in a broad class of quantum systems, including those relevant for quantum thermodynamics and quantum information processing. Future work may explore extensions of the proposed framework to open quantum systems and the incorporation of alternative complexity measures beyond those considered in this study.

\begin{acknowledgments}
 MGA and ROV acknowledge CNPq, CAPES, and FAPEMIG for the financial support provided. TD and ATC acknowledge support from CNPq (Grant No. 441774/2023-7). ATC acknowledges (RAU $N^\circ\,12-2016$-AY-UNA). TD acknowledges support from CNPq (Grant No. 445150/2024-6).
\end{acknowledgments}
\appendix
\renewcommand\thesection{A}
\section{Numerical Evaluation of the Upper Bounds for the Observable Complexity Measures}
\label{app_A}
\addcontentsline{toc}{section}{Appendix A. Numerical Evaluation of the Upper Bounds for the Observable Complexity Measures}

In this appendix, we provide a brief discussion of the theoretical upper bound presented in Eq.~\eqref{eq:bound_C}, which quantifies an asymptotic limitation on the measure associated with the observable's outcomes. These bounds are rigorous and follow from known results in the literature concerning equilibration of closed quantum systems~\cite{meier2025emergence,Reimann2012,short2011equilibration,Short_2012}. The time-averaged observable complexity measure is bounded as
\begin{equation}
\left\langle C(\vec{p}_t) \right\rangle_T \leq \frac{\log r}{2} \sqrt{ \frac{r}{\deff} f(\varepsilon, T) },
\end{equation}
where
\begin{equation}
\label{eq:APP_bound_f}
f(\varepsilon, T) = N(\varepsilon) \left(1 + \frac{8 \log_2(n)}{\varepsilon T} \right),
\end{equation}
and $N(\varepsilon)$ denotes the number of distinct energy gaps within a window of width $\varepsilon$. This expression shows that while the effective dimension $\deff$ plays a fundamental role in the bound, it does not solely determine the time-dependent complexity of the dynamics.


The computation of the bound in Eq.~\eqref{eq:bound_C} requires the evaluation of the function \( f(\varepsilon, T) \), which encodes the fine-grained structure of the energy spectrum through the quantity \( N(\varepsilon) \), the number of distinct energy gaps within a window of width \( \varepsilon \), and incorporates the temporal scale via the factor \( 1/(\varepsilon T) \). This makes explicit that the time-dependent behavior of the observable complexity measure is not dictated solely by the effective dimension \( d_{\mathrm{eff}} \), but rather results from an interplay between spectral properties and dynamical time scales.

Nonetheless, in the asymptotic regime \( T \to \infty \), it becomes possible to choose \( \varepsilon \) such that the function \( f(\varepsilon, T) \) is well-approximated by unity. In this limit, the bound simplifies to
\begin{equation}
\label{bound_app_mean_C}
\left\langle C(\vec{p}_t) \right\rangle_T \lesssim \frac{\log r}{2} \sqrt{ \frac{r}{\deff} },
\end{equation}
revealing a direct inverse square-root dependence on the effective dimension, which highlights the role of quantum state delocalization in suppressing observable complexity at equilibrium.

\begin{figure}[h!]
\centering
\subfloat[\justifying
    Time-averaged observable equilibration complexity measure, \( \langle C(\vec{p}_t) \rangle_T \), for different initial states. The dashed horizontal lines represent the theoretical upper bounds, as given by Eq.~\eqref{eq:bound_C}.
    \label{fig:mean_C_t_APP}
]{\includegraphics[width=7.0cm]{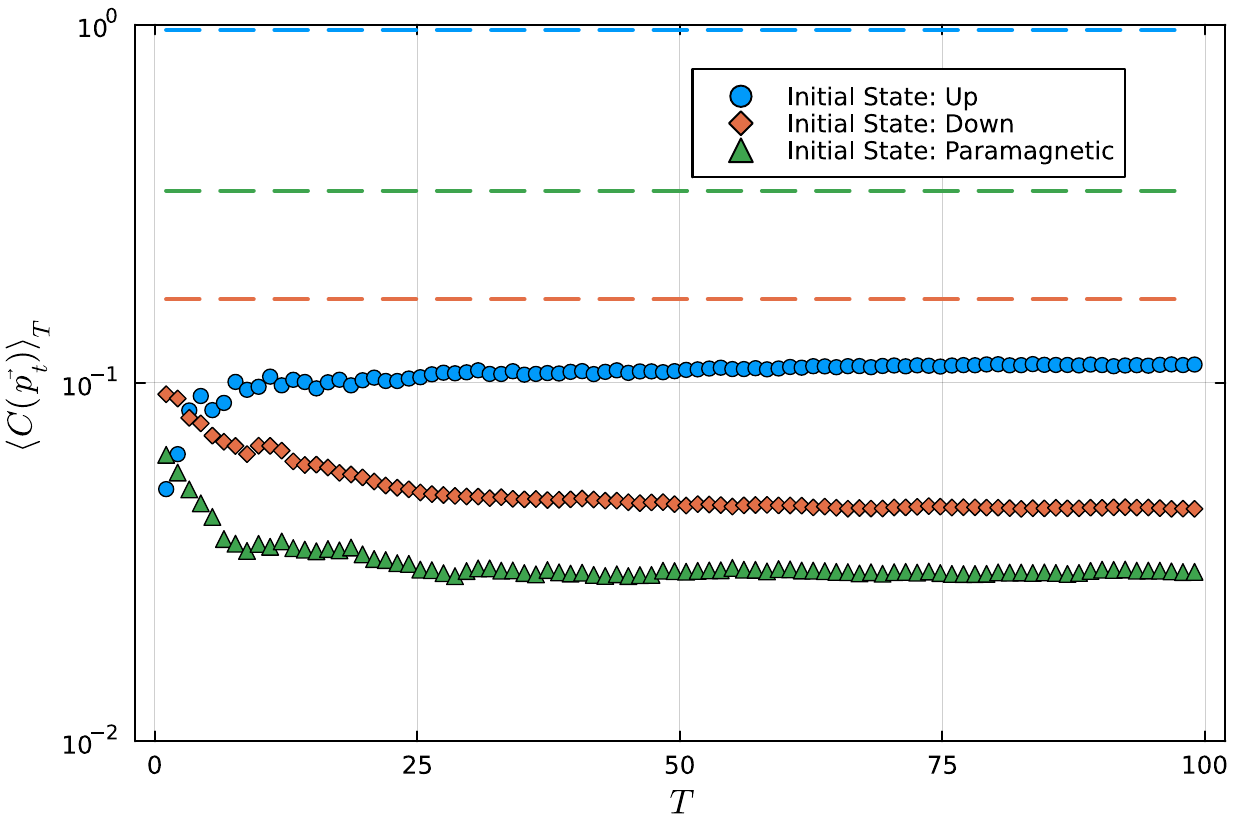}}
\hspace*{0.5cm}
\subfloat[\justifying
    Observable complexity measure \( C(\langle \vec{p}_t \rangle_T) \) computed for the time-averaged distribution. As in panel (a), the dashed horizontal lines indicate the asymptotic bounds from Eq.~\eqref{eq:bound_C}.
    \label{fig:C_of_mean_APP}
]{\includegraphics[width=7.0cm]{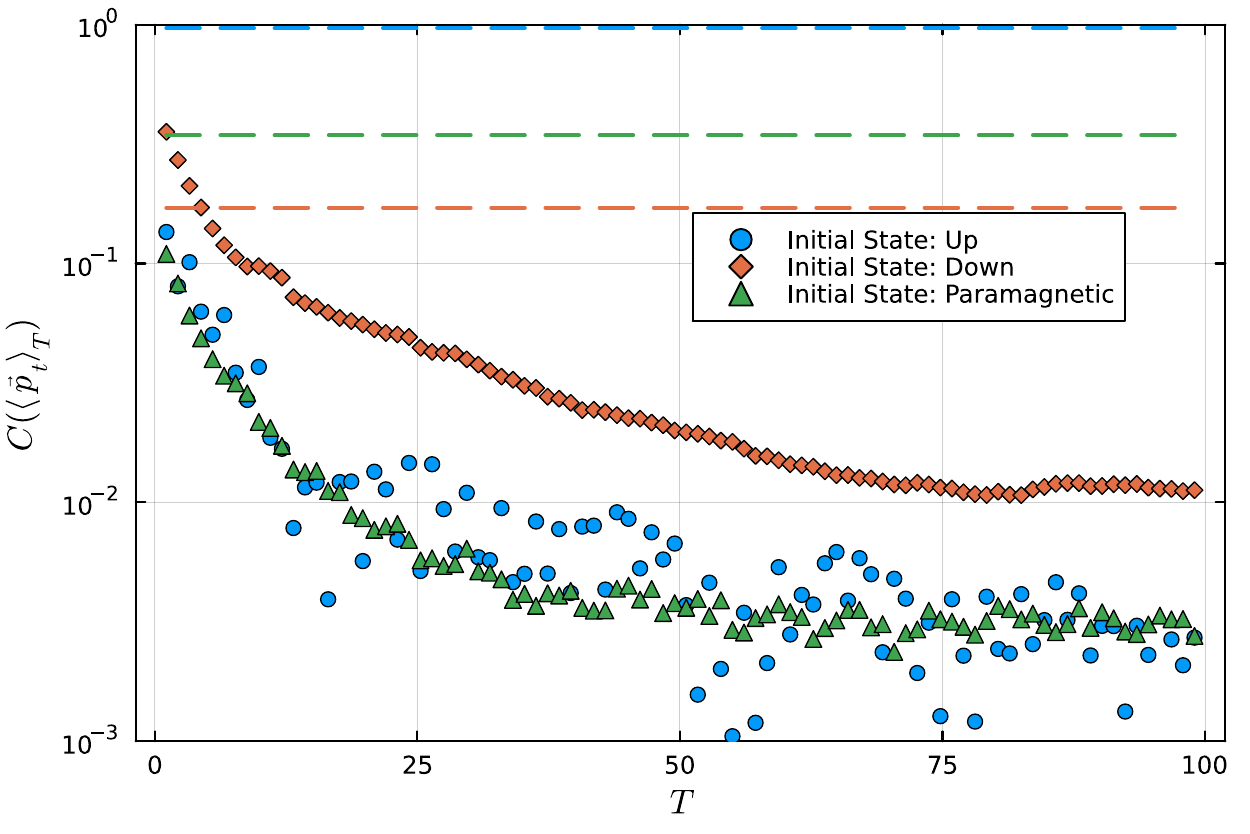}}
\caption{\justifying 
Logarithmic-scale visualization (\( y \)-axis in \(\log_{10}\)) of the observable equilibration complexity for different initial states. Panel (\textbf{a}) shows the time-averaged observable equilibration complexity measure, \( \langle C(\vec{p}_t) \rangle_T \), as a function of the averaging time \( T \). Panel (\textbf{b}) displays the observable complexity computed from the time-averaged probability distribution, \( C(\langle \vec{p}_t \rangle_T) \). In both panels, blue, orange, and green curves correspond to the initial states \texttt{Up}, \texttt{Dw}, and \texttt{Pm}, respectively. The dashed horizontal lines denote the theoretical upper bounds, where the function \( f(\varepsilon, T) \to 1 \) in Eq.~\eqref{eq:bound_C}, yielding the simplified expression \( \frac{\log r}{2} \sqrt{r/\deff} \), for $\deff^{\texttt{Up}}$, $\deff^{\texttt{Dw}}$ and $\deff^{\texttt{Pm}}$.
}
\label{fig:comparison_time_average_APP}
\end{figure}

This asymptotic behavior of the theoretical upper bound is illustrated in Figure~\ref{fig:comparison_time_average_APP}. The Panel~\ref{fig:mean_C_t_APP} displays the time-averaged observable equilibration complexity measure \( \langle C(\vec{p}_t) \rangle_T \) for different initial states. The dashed lines correspond to the theoretical bounds computed via Eq.~\eqref{eq:bound_C}, demonstrating their validity across a range of scenarios. In contrast, Panel~\ref{fig:C_of_mean_APP} presents the observable complexity measure computed from the time-averaged probability distribution, \( C(\langle \vec{p}_t \rangle_T) \), thereby offering a global perspective on how the observable complexity of the measurement outcomes evolves. In both panels, each color denotes a distinct initial state: \texttt{Up} (blue), \texttt{Dw} (orange), and \texttt{Pm} (green). Together, these plots highlight the consistency between numerical data and theoretical expectations, emphasizing the predictive power of the bound and the central role played by the effective dimension in the equilibration of the observable complexity.

It is important to emphasize that the effective dimension \( d_{\mathrm{eff}} \) should not be interpreted as a direct indicator of the complexity of the observable dynamics. As illustrated in Figure~\ref{fig:mean_C_t_APP}, the initial state \texttt{Dw} exhibits a higher effective dimension than \texttt{Pm}, which in turn is greater than that of \texttt{Up}, yet the corresponding values of the time-averaged complexity measure \( \langle C(\vec{p}_t) \rangle_T \) do not follow the same ordering. These results highlight that the bound given in Eq.~\eqref{eq:bound_C} is always satisfied, confirming its theoretical validity, but it is generally not tight. The prefactor \( f(\varepsilon, T) \), which modulates the bound, becomes significant outside the equilibration regime, thereby weakening the informative power of the inequality. Therefore, comparisons between different initial states must be made with caution: the effective dimension alone is insufficient to characterize the observable's dynamical disorder or to predict the saturation of the bound.

As discussed in Section~\ref{sec:complexity_bounds}, Definition~\ref{def: timeaverage_complexity} can be viewed as a particular instance of Definition~\ref{def:OECM}. Within this framework, it satisfies the upper bound presented in Eq.~\eqref{eq:bound_C} of Theorem~\ref{theo: OECM-UB}. As mentioned in Eq. (26), Theorem 3 of Ref. \cite{Short_2012}, the bound for \( C(\langle \vec{p}_t \rangle_T) \) is rigorously valid for all values of \( T \) ``big enough'', that is, its practical relevance depends critically on the behavior of the prefactor \( f(\varepsilon, T) \). In particular, the bound becomes informative only in the regime where \( f(\varepsilon, T) \approx 1 \), which corresponds to timescales associated with equilibration. This regime is characterized by two key conditions: (i) the number of effectively contributing energy gaps \( N(\varepsilon) \) must be moderate, indicating that the spectral resolution is saturated; and (ii) the averaging time must satisfy \( T \gg \frac{8 \log_2(n)}{\varepsilon} \), so that the term \( \frac{1}{\varepsilon T} \) becomes negligible. Under these circumstances, in the same sense of Eq.~\eqref{bound_app_mean_C}, the bound also simplifies to
\begin{equation}
C(\langle \vec{p}_t \rangle_T)\lesssim \frac{\log r}{2} \sqrt{ \frac{r}{\deff} },
\end{equation}
highlighting a scaling inversely proportional to the effective dimension. This behavior is illustrated in Figure~\ref{fig:C_of_mean_APP}, particularly for the state \texttt{Dw}, which exhibits a significant gap between the measured complexity and the theoretical upper bound at short times \cite{Short_2012}. This discrepancy does not indicate a violation of the bound, but rather that the condition \( T \gg \frac{8 \log_2(n)}{\varepsilon} \) is not yet met, and the prefactor \( f(\varepsilon, T) \) remains large. As \( T \) increases, the system approaches the equilibration regime and the bound becomes tighter.

The term \( \frac{8 \log_2(n)}{\varepsilon T} \) presented in Eq.~\eqref{eq:APP_bound_f} arises from bounding time-averaged quantities involving oscillatory terms of the form \( e^{i(E_n - E_m)t} \) \cite{short2011equilibration,Short_2012}. These terms dephase over time, and when \( T \gg 1/\varepsilon \), their contributions tend to cancel out. However, the cancellation is not perfect for finite \( T \), and the residual effect is captured by a correction term inversely proportional to \( \varepsilon T \), multiplied by \( \log_2(n) \), which reflects the number of distinct frequencies. Therefore, the condition \( T \gg \frac{8 \log_2(n)}{\varepsilon} \) ensures that the correction becomes negligible, and the bound approaches its asymptotic, informative regime.

This lack of saturation is expected from the general form of the bound and reflects the difference between typicality-based constraints and detailed dynamical behavior. For this reason, we have chosen to relegate the bound and its verification to this appendix, as it confirms the consistency of the numerical data with theoretical expectations but does not enhance the interpretative analysis of the observed dynamics.

\bibliographystyle{apsrev4-2}
\bibliography{bibfile}

\end{document}